\documentstyle[12pt]{article}
\input{epsf.sty}
\def\reff#1{(\ref{#1})}
\def\ofo{ { {}_2 \! F_1 }}

\newcommand{\beq}{\begin{equation}}
\newcommand{\eeq}[1]{\label{#1}\end{equation}}
\newcommand{\bea}{\begin{eqnarray}}
\newcommand{\eea}[1]{\label{#1}\end{eqnarray}}

\begin{document}
\baselineskip 18pt
\begin{titlepage}
\hfill  NYU-TH-99/09/03

\begin{center}
\hfill
\vskip .4in
{\large\bf Singleton field theory and Flato - Fronsdal dipole equation}
\end{center}
\vskip .4in
\begin{center}
{  Andrei Starinets$  $\footnotemark}
\footnotetext{e-mail: andrei.starinets@physics.nyu.edu}
\vskip .1in
%{\em (a) Theory Division CERN, Ch 1211 Geneva 23, Switzerland}
%\vskip .1in 
{\em  Department of Physics, New York University, 4 Washington Place,
New York, NY 10003, USA}
\end{center}

\vskip .4in
\begin{center}
 {\bf Abstract}
 \end{center}

\begin{quotation}
\noindent

 We study solutions of the equations $(\triangle -\lambda )\phi = 0$ 
and  $(\triangle -\lambda )^2\phi = 0$ in global coordinates
 on the covering space $CAdS_d$ of the  $d$-dimensional
 Anti de-Sitter space subject to various boundary conditions 
 and their connection to the unitary
 irreducible representations of    $\widetilde{SO
}(d-1,2)$. 
The ``vanishing flux'' boundary conditions at spatial infinity lead to the standard quantization scheme for  $CAdS_d$ in which solutions of the second- and the fourth-order equations are equivalent. To include fields realizing the singleton unitary representation in the bulk of  $CAdS_d$ one has to relax the boundary conditions thus allowing for the nontrivial space of solutions of the dipole equation known as the Gupta - Bleuler triplet. We obtain explicit expressions for the modes of the  Gupta - Bleuler triplet and the corresponding two-point function. To avoid negative-energy states one must also introduce an additional constraint in the space of solutions of the dipole equation.
\end{quotation}
\vfill
\end{titlepage}
\eject
\noindent
\section{Introduction}
The purpose of this paper is to obtain explicit solutions in global coordinates to the second- and fourth-order field equations on the $d$-dimensional covering of the Anti de Sitter space, and to establish  their connection to the unitary irreducible representations of the isometry group $\widetilde{SO}(d-1,2)$. 
One may be interested in finding explicit expressions for the modes of the singleton field theory in a view of $AdS/CFT$ correspondence
 \cite{M}, its Lorentzian version \cite{Kraus} and other related issues
\cite{KlebanovWitten}, \cite{singleton_physics}, \cite{Kogan}.
Anti de Sitter space  $AdS_d = SO_0(d-1,2)/SO_0(d-1,1)$ possesses two well-known properties --- rather pathological ones from the physical point of view: it has closed timelike curves and lacks a global Cauchy surface \cite{HE}. The first pathology can be cured by considering covering space $CAdS_d$;
 the second one requires introduction of the special boundary conditions for the solutions of field equations  at spatial infinity \cite{isham}, \cite{BF}.
We are interested therefore in studying $CAdS_d$ harmonics classified according to the representations of the covering group $\widetilde{SO}(d-1,2)$
 of $SO(d-1,2)$.
\paragraph{Representation theory}
In the  literature\footnote{
Mathematical literature provides comprehensive treatment of harmonic analysis on $AdS_d$ (summary and references can be found in \cite{vilenkin} which differs from the physically interesting case of field theory on $CAdS_d$.}, one can find numerous examples of classification of UIRs
 of Lie algebra $so(d-1,2)$ (and corresponding superalgebra), especially for $d=4,5$ \cite{evans}(review and references can be found in \cite{minw}). UIRs of $\widetilde{SO}(d-1,2)$ can be decomposed into the direct sum of UIRs of its maximal  compact
 subgroup, $SO(d-1)\times SO(2)$ and are uniquely characterized by the eigenvalue $\omega$ of the $SO(2)$ generator and by weights of $SO(d-1)$. 
The eigenvalue $\omega$ is always\footnote{Considering only physically interesting UIRs, i.e. those with real positive $\omega$.}
 of the form $\omega=E_0+k$, where $k$ is a nonnegative integer. The condition of unitarity imposes constraint on values of $E_0$,
\beq
E_0\geq E_0^{min},
\eeq{001}
where $E_0^{min}$ depends on the dimension of $CAdS_d$ 
and the type of representation.
UIRs of $\widetilde{SO}(d-1,2)$ exhibit certain unusual (by the standards of the UIRs of compact groups) properties. When the unitarity bound \reff{001}
is saturated,  the number of components in the multiplet is dramatically reduced in comparison with the multiplet characterized by any $E_0 > E_0^{min}$.
More precisely, if we denote states in a given multiplet by $|E_0,l,k>$, where
$l$ is a (set of) quantum numbers of $SO(d-1)$ and $k=0,1,\dots $, then for $E_0=E_0^{min}$ all states  $|E_0^{min},l,k>$ with $k>0$ have zero norm.
 This peculiar UIR (first discussed in a group-theoretical
 approach by Ehrman \cite{Ehrman} and Dirac \cite{Dirac} for $d=4$) is called the  {\it singleton representation}\footnote{Under this name it appeared in Ehrman's paper \cite{Ehrman}. In dimensions $d=5$ and $d=7$ the corresponding UIRs were named ``doubletons''\cite{Gunn}.}. The tensor product of two singleton representations decomposes into an infinite set of massless UIRs \cite{FlatoLett}; hence singletons may be regarded as true fundamental degrees of freedom in 
the  appropriate field theory. This observation served as a powerful motivation for the subsequent development of the singleton field theory in the bulk of 
 $CAdS_4$ \cite{singleton_physics}. Also,
 singleton short multiplets (or supermultiplets, if one considers supersymmetric extensions of $so(d-1,2)$) received much attention in the  '80s in connection with the dimensional reduction of supergravity theories \cite{Gunn},
 \cite{Nicolai}, \cite{sezgin}. Singleton representation in the light of 
  $AdS/CFT$ correspondence is considered in \cite{ferrara1, ferrara2}.
\paragraph{Field theory on $CAdS_d$}

Field theory on four-dimensional\footnote{Generalization to arbitrary $d$ 
is straightforward.
 For explicit solutions of the second-order equation, see \cite{Kraus} and Section 2 of the present paper.}
Anti de-Sitter space
was extensively investigated some time ago by various authors
\cite{isham}, \cite{BF},\cite{fr75} ---\cite{FlFr3}. 
Imposing vanishing flux  boundary conditions (thoroughly described in
\cite{BF}) at the $CAdS_d$ spatial infinity, one finds that 
for $E_0>E_0^{min}$ solutions of the wave equation on $CAdS_d$ are in one-to-one correspondence with the UIRs of  $\widetilde{SO}(d-1,2)$ just like ordinary 
 spherical harmonics on $S^d=SO(d+1)/SO(d)$ are associated with 
UIRs of $SO(d)$.
More precisely, singular 
 points of the differential equation
 $\left(\triangle_{CAdS_d} -\lambda (E_0) \right)\phi =0$ 
in the interval 
\beq
\frac{d-3}{2} < E_0 < \frac{d+1}{2}
\eeq{9080}
 are of the limit-circle  \cite{Coddington} type.
Vanishing flux boundary conditions are chosen in such a way that Cauchy problem on $CAdS_d$ becomes well-defined \cite{isham, BF}. This ultimately gives frequency quantization and establishes connection with   $\widetilde{SO}(d-1,2)$ UIRs. 
Two sets of square-integrable $CAdS_d$ harmonics exist 
for $E_0$ in the interval \reff{9080}.
Outside the interval \reff{9080} singular points are of the limit-point type and the condition of square-integrability leaves only one set of modes for
$E_0 = (d-3)/2$ and $E_0 \geq (d+1)/2$.
The theory is then quantized following the standard scheme \cite{BD}.

\paragraph{Singleton field theory}

The  modes at $E_0 = (d-3)/2$ allowed by the traditional (vanishing flux and square-integrability) boundary conditions
transform according to the $D(\frac{d+1}{2},0)$ 
 UIR of  $\widetilde{SO}(d-1,2)$. Those boundary conditions therefore do not allow realization of the singleton UIR in the bulk of  $CAdS_d$.

However, the second-order equation at  $E_0 = (d-3)/2$ also admits another, non
square-integrable, solution which has logarithmic singularity at the boundary 
of  $CAdS_d$. The logarithmic singularity disappears and the solution 
becomes a polynomial if we impose frequency quantization
$\omega = (d-3)/2+l$. Thus we obtain non square-integrable
 set\footnote{ Non square-integrability here 
does not signal the appearance of the continuous
 spectrum as can be seen clearly by recasting the eigenvalue
 equation into the Schr\"{o}dinger form (see Appendix~\reff{appendix_c}).}
 of the {\it singleton modes} transforming according to 
 $D(\frac{d-3}{2},0)$. The total space of solutions at 
 $E_0 = (d-3)/2$ corresponds to the indecomposable representation 
 $D(\frac{d-3}{2},0)\rightarrow D(\frac{d+1}{2},0)$.
Singleton modes  fall off more slowly at spatial infinity
 than any  solution with $E>E_0^{min}$.
 In this sense they  decouple from the rapidly decreasing
 ``gauge modes'' (solutions bearing 
 $D(\frac{d+1}{2},0)$ labels) 
only at the boundary of $CAdS_d$   
\footnote{
The situation described above  can be compared to other examples provided by harmonic analysis on noncompact groups where
certain members of the {\it discrete} series can not 
be realized as elements of the Hilbert space of square-integrable functions on the corresponding homogeneous space and thus do not appear in the Plancherel formula for the group. Comprehensive analysis for the case of $SO(2,1)$ can be found in Bargmann's 
paper \cite{bargmann}. 
The singleton representation of $\widetilde{SO}(d-1,2)$  is an example of such situation.}.

 ``Singleton modes'' on $CAdS_4$ were introduced and
 studied by C.Fronsdal, M.Flato and collaborators in the series
 of papers \cite{FlFr1} --- \cite{Flato8} (see also 
\cite{singleton_physics} and references therein). Studying the corresponding two-point function, Flato and Fronsdal \cite{FlFr2}
 proposed a  fourth-order wave equation to describe
 a field theory of ``singletons''
 (which was regarded as a gauge theory in the bulk of $AdS_4$)
 and formulated the appropriate Lagrangian formalism. Quantization of the theory including the BRST approach was also developed.
Flato - Fronsdal wave equation is supposed 
to provide the Gupta - Bleuler triplet of ``scalar'', 
``singleton'' and ``gauge'' modes \cite{FlFr2}, \cite{Araki},
\cite{singleton_physics} required for quantization of the theory. The second-order equation then serves as an analog
 of the Lorentz gauge condition.
 Explicit solutions of the fourth-order equation carrying $\widetilde{SO}(d-1,2)$
 UIRs have not been  previously obtained.

In this paper, we study solutions of the second- and the fourth-order 
free field equations on $CAdS_d$ in global coordinates. In Section 2 we
consider modes of the second-order equation subject to vanishing flux
 boundary conditions,
and the singleton modes. We compute the corresponding two-point function ---
first by solving the equation directly and then by using the mode expansion --- and discuss the $E_0\rightarrow (d-3)/2$ limit.
Solutions of the Flato - Fronsdal dipole equation are obtained in Section 3,
modes of the Gupta - Bleuler triplet are identified.
We show that the constraint must be imposed on the space of solutions of the fourth-order equation to eliminate the ``scalar sector'' solutions with the lowest frequency\footnote{Those solutions are not polynomial. The action of energy-lowering operators on them leads to the states with negative energy. }.
We also obtain the singleton two-point function and 
its decomposition into the sum of  Gupta - Bleuler triplet modes.
\section{$CAdS_d$ harmonics and boundary conditions}
In $R^{d+1}$ with metric $\eta_{ij} = (+, -, ..., -, +)$  d-dimensional Anti-de Sitter space $AdS_d$ can be realized as hyperboloid 
\beq
X_0^2 + X_d^2 - \vec{X}^2 \,=\, 1/a^2,
\eeq{1}
where $a=const$ is  the ``inverse radius'' of $AdS_d$.
It is a space of constant curvature 
 $R=d(d-1)a^2$
which is locally characterized by
\beq
 R_{ijkl} \,=\, \frac{R}{12}\left( g_{ik}g_{jl} \,-\, g_{il}g_{jk}\right),
\eeq{2}
The Ricci tensor is proportional to the metric,
\beq
 R_{ij} \,=\, \frac{R}{d} g_{ij} \,=\, (d-1)a^2 g_{ij}.
\eeq{201}
$AdS_d$ can be viewed as a solution of 
Einstein's equations 
\beq
R_{ij} \,-\, \frac{1}{2}g_{ij}R \,=\, -\Lambda g_{ij},
\eeq{202}
with 
$$
\Lambda \,=\, \frac{d-2}{2d}R \,=\, \frac{(d-2)(d-1)}{2}a^2.
$$
It is convenient to use the parametrization

\beq
X^0 \,=\, \frac{\sec{r}}{a}\sin{t},
\eeq{3}

\beq
X^d \,=\, - \frac{\sec{r}}{a}\cos{t},
\eeq{4}

\beq
X^i \,=\, \frac{\tan{r}}{ar}z^i,
\eeq{5}
%\end{document}
where $i=1,\dots , d-1$, $r^2 = \vec{z}\vec{z}$.  We shall use
 $d-1$   ---dimensional spherical coordinates for $z^i$.
The range of $t$ is $[-\pi,\pi)$ for $AdS_d$ and $(-\infty , \infty )$ for its covering space, $CAdS_d$, the range of $r$ is $[0,\pi /2)$, where $r = \pi /2$ corresponds to spatial infinity. The metric can be written as
\beq
ds^2 \,=\, \frac{\sec^2 r }{a^2} \left( dt^2 - dr^2 - \sin^2 r d\Omega_{d-2}^2\right).
\eeq{6}
The  metric on slices $t=const$
 is conformally equivalent to a  half of the sphere $S^{d-1}$ with scalar curvature $R = -\frac{d-2}{d}R_{AdS_d}$:
\beq
ds^2 \mid_{t=const} \,=\, - \frac{\sec^2 r }{a^2} d\Omega_{d-1}^2.
\eeq{7}
 This fact is helpful in discussion of the boundary conditions on $CAdS_d$.

Generators of the $CAdS_d$ isometry group,
\beq
L_{AB} \,=\, X_A \frac{\partial}{\partial X^B} \,-\,   X_B \frac{\partial}{\partial X^A},
\eeq{7a}
$A,B=0,...d$,
can be written in intrinsic coordinates \reff{3} --- \reff{5}  
as
\beq
L_{0d}\,=\,\frac{\partial}{\partial t}, 
\eeq{7b}
\beq
L_{i0}\,=\,- \cos t \sin r \frac{z^i}{r}\frac{\partial}{\partial t}
 - \sin t \left[ \cos r \frac{z^i z^k}{r^2}\frac{\partial}{\partial z^k} + \frac{r}{\sin r}\left( \frac{\partial}{\partial z^i} -  \frac{z^i z^k}{r^2}\frac{\partial}{\partial z^k}\right)\right] , 
\eeq{7bb}
\beq
L_{id}\,=\,- \sin t \sin r \frac{z^i}{r}\frac{\partial}{\partial t} + \cos t \left[ \cos r \frac{z^i z^k}{r^2}\frac{\partial}{\partial z^k} + \frac{r}{\sin r}\left( \frac{\partial}{\partial z^i} -  \frac{z^i z^k}{r^2}\frac{\partial}{\partial z^k}\right)\right] , 
\eeq{7ba}
\beq
L_{ij}\,=\, z^j\frac{\partial}{\partial z^i} \,-\, z^i\frac{\partial}{\partial z^j}.
\eeq{7bc}
Elements of the corresponding $so(d-1,2)$ algebra, $M_{AB}=iL_{AB}$, satisfy
\beq
\left[ M_{AB}, M_{CD}\right] \,=\, i\left( \eta_{BC}M_{AD} \,-\, \eta_{AC}M_{BD}\,-\,  \eta_{BD}M_{AC} \,+\,  \eta_{AD}M_{BC}\right).
\eeq{7c}

The operators  $M_{ik}$ and $M_{0d}$, $i,k=1,...,d-1$,
 generate subalgebra $so(d-1)\otimes u(1)$ of $so(d-1,2)$, $M_{0d}$ being identified as the energy operator on $CAdS_d$. Representations can be built by acting by the energy  raising (lowering) operators,
\beq
M_{k}^{\pm}\,=\,iM_{0k}\,\mp \, M_{kd},
\eeq{7d}
on the lowest energy state satisfying $M_i^- |E_0,j_1,\dots j_{\left[\frac{d-1}{2}\right]} > =0$.
Explicit expressions for $M_{ik}$ in the $d=4$ case (in coordinates $r,t,\theta ,\phi$ )
are
\beq
M_k^{\pm}\,=\, - e^{\mp it}\sin r \frac{z^i}{r}\partial_t \,\mp \, i e^{\mp it}R_i,
\eeq{7e}
where
\beq
R_1\,=\, \sin \theta \cos \varphi \cos r \partial_r \,+\, \frac{\cos \theta \cos \varphi}{\sin r}\partial_{\theta} \,-\, \frac{\sin \varphi }{\sin r \sin \theta}\partial_{\varphi}, 
\eeq{7f}
\beq
R_2\,=\, \sin \theta \sin \varphi \cos r \partial_r \,+\, \frac{\cos \theta \sin \varphi}{\sin r}\partial_{\theta} \,+\, \frac{\cos\varphi }{\sin r \sin \theta}\partial_{\varphi}, 
\eeq{7g}
\beq
R_3\,=\, \cos \theta  \cos r \partial_r \,-\, \frac{\sin \theta }{\sin r}\partial_{\theta}
\eeq{7h}

We consider equation of the form
\beq
\left( \Box_{CAdS_d} - \lambda \right)\phi \, =\, 0,
\eeq{8}
where $\lambda = m^2_0$ is a mass parameter.
Acting on functions, $\Box_{CAdS_d}$ gives
\beq
   \Box f \,=\, -g^{ij} \nabla_j\nabla_i f  \,=\,
-g^{ij}\partial_i\partial_j f + g^{ij}\Gamma^l_{ij}\partial_l f \,=\, -\frac{1}{|g|^{1/2}}\partial_i \left( g^{ij}|g|^{1/2}\partial_j f\right)
\eeq{10}

In coordinates \reff 3 --- \reff 5 the Laplacian on $CAdS_d$ is
\beq
\Box_{CAdS_d} \,=\, -a^2 \cos^2 r \left( \frac{\partial^2}{\partial t^2} -
\frac{\partial^2}{\partial r^2}\right) + (d-2)a^2 \cot r \frac{\partial }{\partial r} + a^2 \cot^2 r \triangle_{S^{d-2}},
\eeq{13}
where $\triangle_{S^{d-2}}$ is the Laplacian on the unit sphere $S^{d-2}$.

A plane-wave solution of
\beq
\Box F \,=\, \lambda F
\eeq{131}
can be written as $F(r,t,\vec{\theta }) = e^{-i \omega t}f(r)Y (\vec{\theta })$,
where $Y (\vec{\theta})$ is an eigenfunction of $\triangle_{S^{d-2}}$,
\beq
\triangle_{S^{d-2}}{\bf Y}({\bf \theta }) \,=\, \Lambda {\bf Y}({\bf \theta }),
\eeq{132}
where $\Lambda = -l(l+d-3)$, $l$ being the highest weight of irreducible representation of $SO(d-1)$.

\begin{figure}[p]
\epsfbox{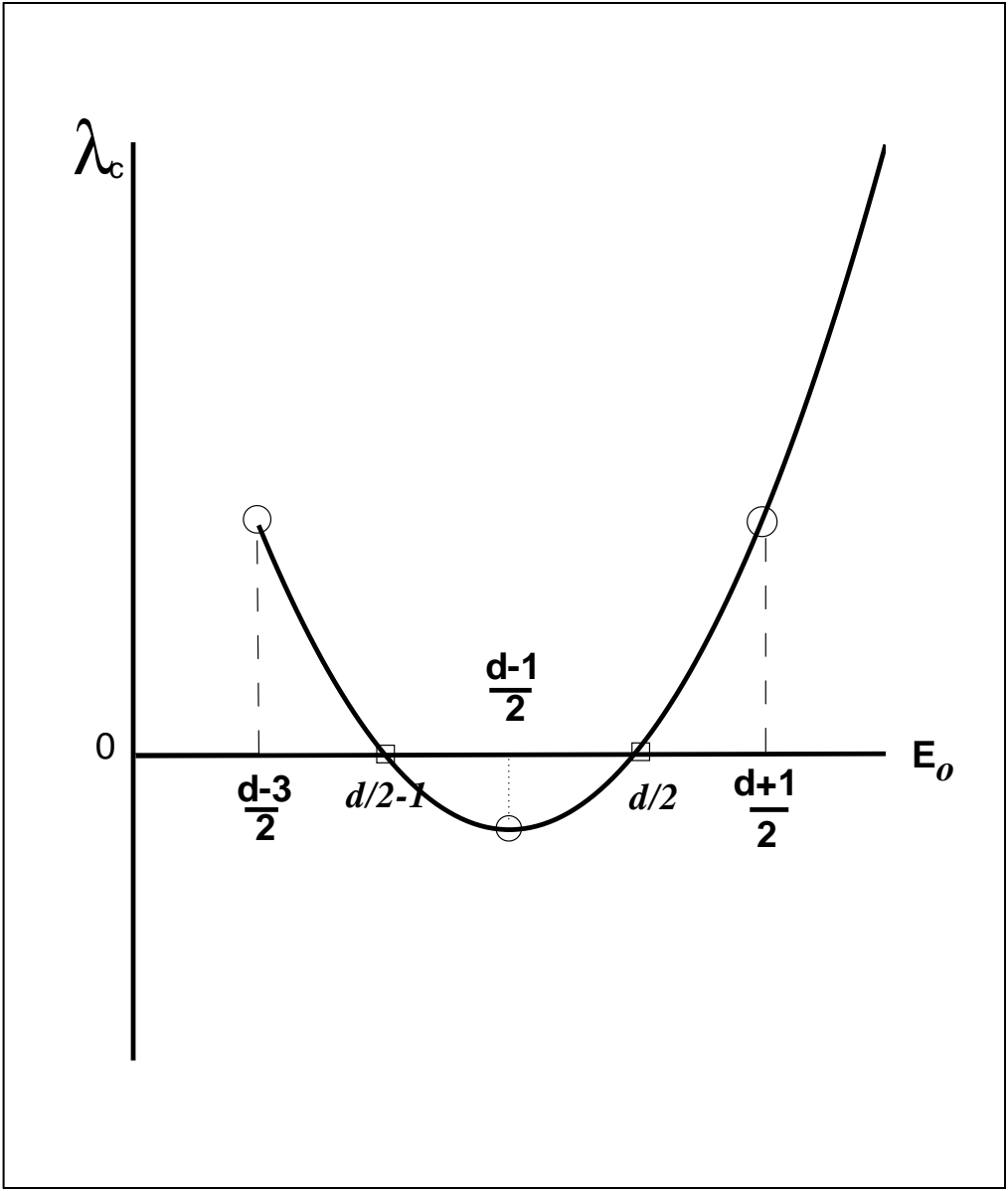}
\caption{Mass squared  $\lambda_c = m^2_c$
 of the conformally coupled scalar
 as a function of $E_0$.
The singleton UIR corresponds to $E_0=\frac{d-3}{2}$. 
Two sets of $CAdS_d$ harmonics exist for $\frac{d-3}{2}<E_0<\frac{d+1}{2}$.
Modes with $d/2-1<E_0<d/2$ have negative mass squared, $m_{BF}^2\leq
 m^2_c < 0$,
 where $m^2_{BF}=-a^2/4$ 
is the Breitenlohner - Freedman
 bound.
}
\label{fig1}
\end{figure}

The equation for $f(r)$ then reads\footnote{ 
It can also be written in the form of the Schr\"{o}dinger equation. In Appendix \reff{appendix_c} we discuss interpretation of the ``singleton modes'' in this picture.}
\beq
Q f \,=\, f''(r) + \frac{2(d-2)}{\sin 2r}f'(r) - \left[ \frac{l(l+d-3)}{\sin^2 r} + \frac{\lambda_p}{a^2\cos^2 r} - \omega^2 \right] f(r) =0.
\eeq{14}
The singular points are $r=0$ and $r=\pi /2$. The roots of the indicial equation at $r=0$ and $r=\pi /2$ are correspondingly
$\alpha^0_1 = l$, $\alpha^0_2 = 3 - d-l$ and 
\beq
\alpha_{1,2}^{\pi /2} = \frac{d-1}{2} \pm \sqrt{\left(  \frac{d-1}{2}\right)^2 +\frac{\lambda}{a^2}}.
\eeq{141}
For a scalar field it is convenient\footnote{Acting on $p$-forms, the second-order Casimir
operator of $so(d-1,2)$ reads $C_2\varphi_p = (E_0-p)(E_0-d+1+p)\varphi_p = m^2_p /a^2\varphi_p$ \cite{ferrara1}.}
 to introduce the parameter $E_0$ such that\footnote{For the scalar field it is sometimes 
 convenient to separate the conformal coupling term from $m_0^2$ and write
$m_0^2 =\frac{d-2}{4(d-1)}R + m_c^2$. Then 
$
m^2_c \,=\,\lambda_c = a^2 \left( E_0 - \frac{d}{2}\right)\left( E_0 - \frac{d}{2}  +1
 \right)$.
This dependence is sketched in Figure~\reff{fig1}. }
\beq
m^2_0 = a^2   E_0 \left( E_0 - d + 1  \right).
\eeq{15}
Then in terms of $E_0$ roots of the indicial equation are
\beq
\alpha_1^{\pi /2} \,=\, E_0,
\eeq{17}
\beq
\alpha_2^{\pi /2} \,=\, d - 1 - E_0.
\eeq{18}
We are looking for solutions of \reff{131} on $CAdS_d$ square-integrable with respect to the metric
\beq
\left( F_1, F_2 \right) \,=\, i \int d^{d-1}x \sqrt{-g} g^{0\nu}\left(
\bar{F_1}\partial_{\nu} F_2 - F_2 \partial_{\nu} \bar{F_1}\right)
\eeq{19}
In coordinates \reff 3 --- \reff 5 this becomes
\beq
\left( F_1, F_2 \right) \,=\, i \int_0^{\pi /2}\left( \frac{\tan r}{a}\right)^{d-2}dr \int d\Omega_{d-2} 
 \left(
\bar{F_1}\partial_{t} F_2 - F_2\partial_{t} \bar{F_1}\right).
\eeq{20}

One can see that for  $r=0$ the solution with index $\alpha_1^0$ is nonsingular and square-integrable for $l > (1-d)/2$ (i.e. for all relevant $l$), the solution with $\alpha_2^0$ is nonsingular for $l\leq 3-d$ and square-integrable for $l < (5-d)/2$. Since we need nonsingular square-integrable solutions for all nonnegative $l$, we choose\footnote{From now on we assume $d>2$. Two-dimensional case is considered in Appendix~\reff{appendix_b}.} solution with $\alpha_1^0$.

For $r = \pi /2$ solutions with index  $\alpha_1^{\pi /2}$ are nonsingular if $E_0\geq 0$ and square-integrable if $E_0 > (d-3)/2$. Solutions with index 
 $\alpha_2^{\pi /2}$ are nonsingular if $E_0\geq d-1$ and square-integrable if $E_0 > (d+1)/2$. One can easily find solutions of \reff{14} with the desired asymptotics. Writing  $f(r)$ as $f(r)=\sin^l{r}\cos^{E_0} r g(r)$, one reduces \reff{14} to a hypergeometric equation in $x=\sin^2 r$:
\beq
x(x-1) g''(x) + g'\left( x(l + E_0 +1) - l - \frac{d-1}{2}\right) +
\frac{(l+E_0)^2 -\omega^2 }{4}g(x) \,=\,0.
\eeq{21}
The solution with correct asymptotics at $r=0$ is\footnote
{Due to the transformation 
property \reff{9000}
 one can also write this solution in the
 form $ \sin^l r \cos^{d-1-E_0} r \ofo \left(\frac{l-E_0 -
\omega + d -1 }{2}, \frac{l-E_0 +\omega +d - 1}{2};
l+\frac{d-1}{2};\sin^2 r \right)$. }
\beq
f(r) \,=\, \sin^l r \cos^{E_0}r  g(r) \,=\, \sin^l r \cos^{E_0}r \, 
\ofo \left( \frac{l+E_0 +\omega}{2}, \frac{l+E_0 -\omega }{2};
l+\frac{d-1}{2};\sin^2 r \right)
\eeq{22}
Consider now the behavior of  \reff{22} at $r\rightarrow \pi /2$. Using a transformation property \reff{999} of hypergeometric functions 
we can conveniently express $f(r)$ as
\begin{eqnarray}
f(r) &=& C_1 \sin^l r \cos^{E_0} r \ofo \left( \frac{l+E_0 +\omega}{2}, \frac{l+E_0 -\omega }{2};
E_0 -\frac{d-3}{2};\cos^2 r \right)\nonumber  \\
&+&C_2 \sin^l r \cos^{d-1-E_0}r \, \ofo \left( \frac{l-E_0-\omega +d-1}{2},
 \frac{l+E_0 +\omega  +d-1}{2};
\frac{d+1}{2} - E_0;\cos^2 r \right), \nonumber
%\label{999}
\end{eqnarray}
where
\beq
C_1 \,=\, \frac{\Gamma\left( l +\frac{d-1}{2}\right)
\Gamma\left( \frac{d-1}{2} -E_0\right)}{\Gamma\left(
\frac{l+d-1-E_0 -\omega }{2} \right)\Gamma\left(
\frac{l+d-1-E_0 +\omega }{2} \right)},
\,  \, \, \,
C_2 \,=\, \frac{\Gamma\left( l +\frac{d-1}{2}\right)
\Gamma\left(E_0 -  \frac{d-1}{2} \right)}{\Gamma\left( 
\frac{l+E_0 +\omega }{2} \right)\Gamma\left( 
\frac{l+E_0 -\omega }{2} \right)}.
\eeq{251}

\subsection{Solution in the interval $\frac{d-3}{2} < E_0 < \frac{d+1}{2}$}
In the range 
\beq
\frac{d-3}{2} < E_0 < \frac{d+1}{2}
\eeq{23}
 (for $d>1$) the singular point 
$r=\pi /2$ is of the limit-circle type (asymptotics with $\alpha_1^{\pi /2}$
and $\alpha_2^{\pi /2}$
 are equally admissible) so we need to specify boundary conditions.  

As mentioned in the Introduction (and discussed in detail in  \cite{BF}), the Cauchy problem on $CAdS_d$ is ill-defined since time development of fields
 can be affected by the information crossing spatial infinity $r =\pi/2$ in finite time. To make evolution predictable, we have to impose the condition that the flux through the boundary $r =\pi /2$ vanishes. Calculation of the flux proceeds exactly as in
 the four-dimensional case \cite{BF}  and one finds that the requirement of vanishing flux is equivalent to setting either $C_1$ or $C_2$ in \reff{22} to zero. 

Another way to describe these boundary conditions is to consider the conformal map 
 \reff 7 from $CAdS_d$ to the half of $S^{d-1}$. The boundary of $CAdS_d$ is mapped into the equator of $S^{d-1}$. Fields transform according to 
$$
\Phi \,=\, \cos^{1-d/2}{r} \, \,   \Phi_{CAdS_d}.
$$
Near the equator of $S^{d-1}$ one has then 
$$
f(r) \,\sim \, C_1 \cos^{E_0 +1 -\frac{d}{2}}{r} \,+\, C_2 \cos^{\frac{d}{2} -E_0}{r} +\mbox{higher powers of }\cos{r}
$$
At $E_0 = d/2 -1$ (massless scalar field) this becomes simply
$$
f(r) \,\sim \, C_1  \,+\, C_2 \cos{r} +\mbox{higher powers of }\cos{r}
$$
Thus conditions $C_1=0$ or $C_2=0$ correspond to ordinary Dirichlet or Neumann boundary conditions, respectively.

We have therefore two sets of modes on $CAdS_d$ corresponding to ``Dirichlet'' ($C_1 = 0$) or ``Neumann'' ($C_2=0$) boundary conditions.
\paragraph{Dirichlet boundary condition}
Condition  $C_1=0$ leads to quantization of $\omega$: $\omega_k = d-1-E_0 +l +2k$, where $k=0,1,\dots$. The second coefficient becomes
$$
C_2 (\omega_k )\,=\,  \Gamma\left( l +\frac{d-1}{2}\right)
\Gamma\left(E_0 -  \frac{d-1}{2} \right) / \Gamma\left( 
\frac{d-1}{2}+l+k \right)\Gamma\left( 
E_0 - \frac{d-1}{2}-k \right).
$$
The solution is
\begin{eqnarray}
 f(r) & = & \sin^l r \cos^{d-1-E_0} r \ofo \left( -k,
 d-1-E_0 +l +k;
l+\frac{d-1}{2};\sin^2 r \right) \nonumber \\  & = &
\sin^l r \cos^{d-1-E_0}{r}\frac{k!}{\left( l+\frac{d-1}{2}\right)_k }
  P_k^{(l+\frac{d-3}{2},
\frac{d-1}{2}-E_0) } \left( \cos 2r \right),
\end{eqnarray}
where $(a)_k = \Gamma(a+k)/\Gamma(a)$ is the Pochhammer's symbol.
Thus, the first set of solutions\footnote{Normalized with respect to the metric \reff{20}.} of equation \reff{131} is
\beq
F^D_{E_0,l,k}(r,t,{\bf \theta } ) \,=\, C_{E_0,l,k} e^{- i \omega_k t} f_k^- (r){\bf Y}({\bf \theta })  ,
\eeq{24}
where
\beq
f_{E_0,l,k}^D (r) \,=\, \sin^l r \cos^{d-1-E_0} r  P_k^{(l+\frac{d-3}{2},
 \frac{d-1}{2}-E_0) } \left( \cos 2r \right),
\eeq{25}
\beq
C_{E_0,l,k}^2 \,=\, \frac{a^{d-2}\Gamma \left(d-1- E_0 + l +k \right) k!}
{\Gamma \left(  \frac{d-1}{2}+ l +k \right)
\Gamma\left(  \frac{d+1}{2}-E_0 +k \right)},
\eeq{25177}
\beq
\omega_k \,=\, d - 1 - E_0 + l + 2k, \; \; \; \; k=0,1,2,\dots .
\eeq{252}
\paragraph{Neumann boundary condition}
 Condition  $C_2=0$ gives frequency quantization $\omega_k
 = E_0 + l +2k$,
$k=0,1,\dots$. Coefficient $C_1$ is  
$$
C_1 (\omega_k )\,=\,  \Gamma\left( l +\frac{d-1}{2}\right)
\Gamma\left( \frac{d-1}{2}-E_0 \right) / \Gamma\left( 
\frac{d-1}{2}+l+k \right)\Gamma\left( 
\frac{d-1}{2}- E_0 -k \right).
$$
The second set of solutions is therefore
\beq
F^N_{E_0,l,k}(r,t, \theta  ) \,=\, C_{E_0,l,k} e^{- i \omega_k t} f_k^+ (r){\bf Y}({\bf \theta }) ,
\eeq{811}
where
\beq
f_{E_0,l,k}^N (r) \,=\, \sin^l r \cos^{E_0} r  \, \, P_k^{(l+\frac{d-3}{2},E_0 - \frac{d-1}{2})} \left( \cos 2r \right),
\eeq{812}
\beq
C_{E_0,l,k}^2 \,=\, \frac{a^{d-2}\Gamma \left( E_0 + l +k \right) k!}{\Gamma \left(  \frac{d-1}{2}+ l +k \right)\Gamma\left( E_0 -  \frac{d-3}{2} +k \right)},
\eeq{813}
\beq
\omega_k \,=\, E_0 + l + 2k, \; \; \; \; k=0,1,2,\dots .
\eeq{814}
\paragraph{Massless modes}
Massless conformally coupled scalar modes correspond to $E_0=d/2$ or $E_0=d/2-1$ and  are given by $F(r,t,{\bf \theta })=f_k(r)e^{-i \omega_k t} {\bf Y}({\bf \theta } )$, where
for $E_0=d/2$ $
\omega_k = d/2 + l + 2k,
$
\beq
f_k^{E_0=d/2}(r) \,=\, A \sin^l r \cos^{\frac{d}{2}-1}r C^{l+\frac{d}{2}-1}_{2k+1}(\cos r ),
\eeq{91}
where $C^{\alpha}_k$ are Gegenbauer polynomials and 
$$
A^2  \,=\, a^{d-2}\Gamma (k+3/2) k! / \pi \left( l+d/2 +1\right)_{k+1} \left( l+d/2 -1\right)_{k+1/2}.
$$
for $E_0=d/2-1$ we have 
$
\omega_k = d/2 -1 + l + 2k,
$
\beq
f_k^{E_0=d/2-1}(r) \,=\, B \sin^l r \cos^{\frac{d}{2}-1}r C^{l+\frac{d}{2}-1}_{2k}(\cos r ),
\eeq{91a}
where 
$
B^2\,=\, a^{d-2}\Gamma (k+1/2) k! / \pi
 \left( l+d/2 +1\right)_{k+1} \left( l+d/2 -1\right)_{k+1/2}.
$

\subsection{Solution outside the interval  $\frac{d-3}{2} < E_0 < \frac{d+1}{2}$}  
Outside the interval  $\frac{d-3}{2} < E_0 < \frac{d+1}{2}$ singular point
 $r=\pi /2$ of the differential equation \reff{14}
is of the limit-point type (boundary condition becomes the condition of square-integrability on $CAdS_d$). This means that for $E_0<\frac{d-3}{2}$ ( $E_0>\frac{d+1}{2}$) coefficient $C_1$ ($C_2$) must be set to zero and, correspondingly, only the Dirichlet set (Neumann set   ) is acceptable. We note that the 
 $E_0<\frac{d-3}{2}$ Dirichlet set is equivalent to the $E_0>\frac{d+1}{2}$ Neumann set and thus for $m_0^2/a^2 > m_{BF}^2/a^2 + 1 = - (d-1)^2/4 + 1$ there exists only one set of normalizable solutions on $CAdS_d$.    

\subsection{Solution for $E_0 = \frac{d-1}{2}+m, m\in Z$} 
When the difference between the roots \reff{17} -- \reff{18} of the indicial equation 
becomes an integer number, i.e. when 
\beq
E_0 \,=\, \frac{d-1}{2} \,+\, m, \hspace{0.5cm} m\in Z,
\eeq{444}
we may expect logarithmic terms in the solution of equation \reff{14}.
For example, when $E_0 = (d-1)/2$, \reff{22} degenerates to 
\begin{eqnarray}
f(r) &=& \frac{\Gamma \left( l +\frac{d-1}{2}\right)}{\Gamma \left( 
a  \right) \Gamma \left( b  \right)}\sin^l r \cos^{\frac{d-1}{2}}r
\sum_{n=0}^{\infty} \frac{(a)_n(b)_n}{(n!)^2} \nonumber \\
&\cdot& \left(   2 \psi (n+1) - 
\psi (a+n) -\psi (b+n) -\log (\cos^2 r )\right) \cos^{2n} r ,
\label{plo}
\end{eqnarray}
where 
$$
a,b \,=\, \frac{1}{2}\left( l +\frac{d-1}{2} \pm \omega \right).
$$ 
One can impose the same boundary conditions as in the non-degenerate case by continuity\footnote{ Alternatively, we may demand the absence of the logarithmic singularity in the solution and set $b=-k$, $k=0,1,\dots$ which gives the same result.
}
 in $E_0$. With the frequency given by \reff{252} or \reff{814}, \reff{plo} gives
\beq
F_k^{E_0=\frac{d-1}{2}} (r,{\bf \theta }) \,=\, a^{\frac{2-d}{2}}
e^{-i\omega_k t}\sin^l r \cos^{\frac{d-1}{2}} r  \, \, P_k^{(l+\frac{d-3}{2},
0)} \left( \cos 2r \right){\bf Y}({\bf \theta }) .
\eeq{81297}
where
$
\omega_k \,=\, \frac{d-1}{2} \,+\, l +2k.
$
Therefore, at $E_0 = (d-1)/2$ two sets of modes \reff{25} --\reff{812} merge into one, normalizable with respect to the metric \reff{20}.
Logarithms do not appear in \reff{81297}. The argument can be repeated for any $m\in Z$ using formulas \reff{10004} - \reff{10005}. We conclude that $E_0 = \frac{d-1}{2}+m, m\in Z$ is {\it not} a special case and it gives the same Dirichlet or Neumann sets of solutions which were discussed in the previous subsections.

\subsection{Singleton modes}
From our analysis of the differential equation \reff{131} it follows that at $E_0=(d-3)/2$ the only set of square-integrable modes is the Dirichlet set  \reff{24} -- \reff{252} (equivalent to the Neumann set \reff{811} -- \reff{814}
 with $E_0=(d+1)/2$ corresponding to $so(d-1,2)$ UIR $D(\frac{d+1}{2},0)$). We observe therefore that the singleton UIR $D(\frac{d-3}{2},0)$ is not realized on ${\cal L}^2(CAdS_d)$.

 It is known, however, that at  $E_0=(d-3)/2$ the representation becomes indecomposable $D(\frac{d-3}{2},0)\rightarrow D(\frac{d+1}{2},0)$ ($\frac{d-3}{2}$ ``leaking'' into  $\frac{d+1}{2}$). For example, acting on the ground state $|\frac{d-3}{2},0,0>$ (satisfying $M^-_i |\frac{d-3}{2},0,0> =0$) the energy-raising operator $M^+_i$ gives the state $|\frac{d-3}{2},1,0>$ but the action of $M^-_i$ on  $|\frac{d-3}{2},1,0>$ does not lead back to $|\frac{d-3}{2},0,0>$ being instead proportional to  $|\frac{d+1}{2},0,0>$,
the ground state of $D(\frac{d+1}{2},0)$. We have therefore a vector space $V$ of states with $E_0=(d-3)/2, k=0$ (singleton or ``physical'' modes) and $E_0=(d+1)/2$ (``gauge'' modes) in which states with  $E_0=(d-3)/2, k=0$ do not form an invariant subspace. The Hilbert space for the singleton UIR is a quotient ${\cal H}=V/V_{gauge}$.

It is possible to realize this description in terms of the fields in the bulk of $CAdS_d$ satisfying equation \reff{131}. In addition to the Dirichlet set
 \reff{24} -- \reff{252} with $E_0=(d-3)/2$ consider the Neumann set  \reff{811} -- \reff{814}  with 
 $E_0\rightarrow (d-3)/2$. 
The limit is 
\begin{eqnarray}
& \, & F^N_{\frac{d-3}{2},l,0}\left( r,t,\theta\right) = 0, \\
&\, &  F^N_{\frac{d-3}{2},l,k}\left( r,t,\theta\right) = \sqrt{\frac{ a^{d-2} k}
{\frac{d-3}{2}+l+k}}
e^{-i \omega_k t}\sin^l r \cos^{\frac{d-3}{2}} r  \, \, P_k^{(l+\frac{d-3}{2},
-1)} \left( \cos 2r \right){\bf Y}( \theta ),
\label{limsin}
\end{eqnarray}
where $\omega_k =  \frac{d-3}{2} \,+\, l +2k$, $k=1,2,\dots$.
Using the identity 
\beq
P_{k+1}^{(\alpha ,-1)}\left( \cos 2r \right) \,=\, \left(\frac{\alpha }{k+1}+1\right) \cos^2 r 
P_{k}^{(\alpha ,1)}\left( \cos 2r \right)
\eeq{90807a}
one can see that the rest of the $E_0=(d-3)/2$  Neumann modes 
(with $k\geq 1$) are equivalent to the standard Dirichlet set  \reff{24} -- \reff{252} with  
 $E_0=(d-3)/2$, $k\geq 0$ (or the standard Neumann set  \reff{811} -- \reff{814} with  $E_0=(d+1)/2$, $k\geq 0$),
\beq
F^{gauge}_{l,k} (r,t,\theta ) \,=\,  a^{\frac{d-2}{2}}\sqrt{\frac{\frac{d-1}{2}+ l+k}{k+1}    }  e^{-i(\frac{d+1}{2}+l+2 k) t}\sin^l r \cos^{\frac{d+1}{2}} r  P_k^{(l+\frac{d-3}{2},1)}(\cos 2r ) 
{\bf Y}({\bf \theta }), \; \; k\geq 0.
\eeq{988a}
Thus taking the limit shows that for $E_0=(d-3)/2$ the only set satisfying the equation \reff{131} and Breitenlohner - Freedman boundary conditions is the Dirichlet set.

{\it Relaxing}  boundary conditions  we discover the singleton modes. Indeed, the solution of \reff{131} at $E_0=(d-3)/2$ is given by 
\begin{eqnarray}
f(r) &=&
\frac{\Gamma \left( a+b+1 \right)}{\Gamma \left( 
a +1 \right) \Gamma \left( b +1 \right)}\sin^l r \cos^{\frac{d-3}{2}}r \nonumber \\ &+&
\frac{\Gamma \left( a+b+1 \right)}{\Gamma \left( 
a  \right) \Gamma \left( b  \right)}
\sin^l r \cos^{\frac{d+1}{2}}r 
\sum_{n=0}^{\infty} \frac{(a+1)_n(b+1)_n}{n!(n+1)!} \nonumber \\
&\cdot& \left( \log (\cos^2 r )-\psi (n+2)-\psi (n+1)+
\psi (a+n+1) +\psi (b+n+1) \right) \cos^{2n}r,
\label{polp}
\end{eqnarray}
where 
$$
a,b \,=\, \frac{1}{2}\left( l +\frac{d-3}{2} \pm \omega \right).
$$ 
Instead of imposing vanishing flux boundary conditions (which would require the first term  in \reff{polp} to vanish and lead to \reff{988a}) we require\footnote{Alternatively, we may demand the absence of logarithmic singularity in the solution which gives the same result.} that $\omega_k$ be equal to $\omega_k \,=\, \frac{d-3}{2} \,+\, l +2k$. This gives
\beq
F^{singleton}_l (r,t, \theta ) \,=\, e^{-i (\frac{d-3}{2}+l) t}\sin^l r \cos^{\frac{d-3}{2}} r  
{\bf Y}( \theta  ) 
\eeq{8199}
for $k=0$ and the set \reff{988a} for $k>0$. 
The singleton modes
 {\it are not square-integrable\footnote{Note that the singleton modes do not satisfy vanishing flux boundary conditions on $CAdS_d$.}} with respect to \reff{20}.
Singleton modes \reff{8199} and gauge modes \reff{988a} form a vector space $V$ which can be equipped with the scalar product
\beq
\left( F_k,F_{k'} \right) \,=\, i\lim_{E_0\rightarrow \frac{d-3}{2}}
 \left( E_0 -\frac{d-3}{2}\right) \int_0^{\pi /2}\left( \frac{\tan r}{a}\right)^{d-2}dr \int d\Omega_{d-2} 
 \left(
\bar{F_k}\partial_{t} F_{k'} - F_{k'}\partial_{t}\bar{ F_{k}}\right)
\eeq{2091}
We observe that $(F_{l,k}^{gauge},F_{l,k}^{gauge})=0$ for all
 $k\geq 0$ and $(F^{singleton}_l,F_l^{singleton})=a^{d-2}\left(l+
\frac{d-3}{2}\right)$.
The singleton modes belong to the Hilbert space  ${\cal H}=V/V_{gauge}$   with the scalar product \reff{2091}.
Singleton modes have a more singular behavior $(f\sim \cos^{\frac{d-3}{2}}r)$
as $r \rightarrow \pi /2$ than the gauge modes $(f\sim \cos^{\frac{d+1}{2}}r)$
and so at the boundary gauge modes ``decouple''. In this sense singletons ``live on the boundary'' of $CAdS_d$.

\subsection{Two-point function}
In this subsection we shall determine the two-point function,
\beq
D^{D,N}(x,x')\,=\, <0|\hat{\Phi}^{D,N}(x)\hat{\Phi}^{D,N}(x')|0>,
\eeq{1tpf}
where
\beq
\hat{\Phi}^{D,N}(x)
\,=\, \sum_{i} \left\{ \phi^{D,N}_i a_i \,+\, \bar{ \phi}^{D,N}_i
 a_i^*\right\},
\eeq{2tpf} 
and $\left\{  \phi^{D,N}_i\right\}$ is 
a complete set (Dirichlet or Neumann ) of the solutions of \reff{131} on $CAdS_d$. We shall do this calculation twice - first by generalizing approach of
 \cite{FlFr2} to $d$ dimensions and then by evaluating the sum 
\beq
D^{D,N}(x,x')\,=\,\sum_i \phi_i^{D,N}(x)\bar{\phi}_i^{D,N} (x')
\eeq{3tpf}
explicitly. The second method has an advantage of providing explicit normalization for $D(x,x')$ determined by the boundary conditions imposed on the fields $\phi_i(x)$.

Due to the $SO(d-1,2)$ invariance,  $D(x,x')$ is a function of $Z=a^2 X \cdot X'$ \cite{FlFr2}. In global coordinates \reff{3} - \reff{5} $Z$ reads
\beq
Z\,=\, \sec r \sec r' \left( \cos (t-t') - \frac{\sin r \sin r'}{rr'}z_iz_i' \right) .
\eeq{4tpf}
The two-point function $D(Z)$ satisfies \reff{131},
\beq
\left[ \left( 1 - Z^2\right) \partial^2_{ZZ} - d Z \partial_Z + E_0 \left( E_0 - d +1 \right) \right] D(Z) = 0,
\eeq{5tpf}
general solution of which can be written as
\beq
 D(Z) \,=\, C_1 (E_0,d) \; \;  \ofo \left( 
 \frac{d-1-E_0 }{2},\frac{E_0}{2};\frac{1}{2} ;  Z^2 \right)
\,+\, C_2(E_0,d) \; \;  Z \; \; \ofo \left( 
 \frac{d-E_0 }{2},\frac{E_0+1}{2};\frac{3}{2} ;  Z^2 \right).
\eeq{6tpf}
Two independent solutions correspond to Dirichlet and Neumann sets of modes.
To make connection with \cite{FlFr2} it is convenient to rewrite \reff{5tpf} --\reff{6tpf} in terms of $\xi = 1/Z$:
\beq
\left[ \xi^2 (\xi^2 - 1 ) \partial^2_{\xi \xi} + \xi \left( 2 \xi^2 + d -2 \right) \partial_{\xi }+ E_0 \left( E_0 - d + 1\right) \right] D(\xi ) = 0,
\eeq{7tpf}
\begin{eqnarray}
D(Z) &=& C_1 (E_0,d) \;  Z^{E_0+1-d} \; \; \ofo \left( 
 \frac{d-1-E_0 }{2},\frac{d-E_0}{2};\frac{d+1}{2}-E_0 ;  \frac{1}{Z^2} \right)
\nonumber \\ 
& +&  C_2 (E_0,d)\;  Z^{-E_0}\; \;  \ofo \left( 
 \frac{E_0 + 1 }{2},\frac{E_0}{2};\frac{3-d}{2}+E_0 ;  \frac{1}{Z^2} \right) .
\label{8tpf}
\end{eqnarray}
%For $d=4$, $E_0=1/2$ we recover the result of \cite{FlFr2}:
%\beq
%D^D(Z) = C_1   Z^{-5/2} \; \;  \ofo \left( 
% \frac{5 }{4},\frac{7}{4};2 ;\frac{1}{Z^2} \right),
%\eeq{9tpf}
%\beq
%D^N(Z) = C_2  Z^{-1/2}\; \;  \ofo \left( 
% \frac{3}{4},\frac{1}{4};0;\frac{1}{Z^2} \right),
%\eeq{10tpf}
Now we shall calculate $D^D(Z)$ and  $D^N(Z)$ using \reff{2tpf} and explicit solutions found in this section. Putting for simplicity $x'=0$ we have $Z=\sec r \cos t$. Then 
\beq
D^D(r,t) = \frac{\Gamma \left( \frac{d-1}{2}+1\right) }{(d-1)\pi^{\frac{d-1}{2}}}\cos^{d-1-E_0}r  \, e^{- i t (d-1-E_0)}\sum_{k=0}^{\infty} C^2_{E_0,0,k}
\frac{ \left(\frac{d-1}{2}\right)_k   }{k!} \, P_k^{(l + \frac{d-3}{2},\frac{d-1}{2}-E_0)}(\cos 2r )e^{-i2kt}.
\eeq{11tpf}
Performing the sum with the help of \cite{prudnikov} we get
\beq
D^D(Z) = \frac{a^{d-2}}{2^{d-E_0}\pi^{\frac{d-1}{2}}}
\frac{\Gamma \left( d-1-E_0\right)}
{\Gamma\left( \frac{d+1}{2}-E_0\right)}
 Z^{E_0-d+1}
\ofo \left( \frac{d-1-E_0 }{2},
 \frac{d-E_0  }{2};
 \frac{d+1}{2} - E_0 ; \frac{1}{Z^2} \right).
\eeq{12tpf}
For the Neumann set we have
\beq
D^N(Z) = \frac{a^{d-2}}{2^{E_0+1}\pi^{\frac{d-1}{2}}}
\frac{\Gamma \left( E_0\right)}
{\Gamma\left(E_0 - \frac{d-3}{2}\right)}
 Z^{-E_0}
\ofo \left( \frac{E_0 }{2},
 \frac{E_0 +1  }{2};
 E_0 - \frac{d-3}{2} ; \frac{1}{Z^2} \right).
\eeq{13tpf}
Explicit expressions for $C_1$ and  $C_2$
allow us to consider the limit $E_0\rightarrow (d-3)/2$:
\beq
\lim_{E_0\rightarrow \frac{d-3}{2}}D^D(Z) \,=\, D^D_{\frac{d-3}{2}} (Z) \,=\, \frac{a^{d-2}\Gamma (\frac{d+1}{2})}{(2\pi)^{4\frac{d-1}{2}}} \;  Z^{-\frac{d+1}{2}} 
\ofo \left( \frac{d+1 }{4},
 \frac{d+3}{4};
 2 ; \frac{1}{Z^2} \right),
\eeq{14tpf}
\beq
\lim_{E_0\rightarrow \frac{d-3}{2}}D^N(Z) \,=\,  D^D_{\frac{d-3}{2}} (Z). 
\eeq{15tpf}
Thus the limit of the Neumann two-point function is simply equal to the Dirichlet  two-point function at $E_0=(d-3)/2$. This result is obviously consistent with the description of the limit     \reff{limsin} of the Neumann set satisfying vanishing flux  boundary conditions.

Relaxing the  boundary conditions and solving \reff{7tpf} {\it directly at
} $E_0=(d-3)/2$ we obtain
\beq
D(Z) \,=\, C_1 w_1 (Z) \,+\, C_2 w_2 (Z),
\eeq{16tpf}
where
\beq
w_1 \,=\, Z^{-\frac{d+1}{2}}\,  \ofo \left( \frac{d+1 }{4},
 \frac{d+3}{4};
 2 ; \frac{1}{Z^2} \right),
\eeq{17tpf}
\begin{eqnarray}
w_2 &=& -  Z^{-\frac{d+1}{2}} \log{Z^2}\,  \ofo \left( \frac{d+1 }{4},
 \frac{d+3}{4};
 2 ; \frac{1}{Z^2} \right)
 +
 Z^{-\frac{d+1}{2}}\, \sum_{n=1}^{\infty} Z^{-2n} \; \frac{ \left(\frac{d+1}{4}\right)_n \left(\frac{d+3}{4}\right)_n }{(2)_n n!}\nonumber \\ 
&\, & \left[ \psi \left( \frac{d+1}{4}\right) + \psi \left( \frac{d+3}{4}\right) - \psi \left( \frac{d+1}{4}\right) -  \psi \left(\frac{d+3}{4}\right) +\psi (2)-\psi (2 + n)+\psi (1) - \psi (n+1) \right] \nonumber \\ &+&
  Z^{-\frac{d-3}{2}}\, \frac{16}{(d-1)(d-3)}.
\label{18tpf}
\end{eqnarray}
with coefficients $C_1$ and $C_2$ remaining undetermined.
The solution \reff{14tpf} is recovered by putting $C_2=0$ which is equivalent to imposing vanishing flux boundary conditions. Another possibility is to relax vanishing flux boundary conditions and introduce the singleton modes \reff{8199}  propagating in the bulk \cite{FlFr2}.
A fourth-order differential equation was proposed in \cite{FlFr2} to describe
the  singleton field theory in the bulk of $CAdS_4$. The original motivation in \cite{FlFr2} was to find an equation for the singleton two-point function 
solution of which would be free from logarithmic singularities. Even though this goal cannot be achieved globally\footnote{The two-point function $ \ofo \left( \frac{1}{4}, \frac{3}{4};1;\frac{1}{Z^2} \right)$, obtained in 
 \cite{FlFr2} as the solution of the fourth-order
 equation avoids logarithmic singularity similar to that one in \reff{18tpf} at the {\it boundary} of  $CAdS_4$ but instead it acquires logarithmic behavior near the origin.}
 on  $CAdS_d$, the proposed equation provides an interesting opportunity for quantization of the singleton modes in the bulk via Gupta - Bleuler triplet technique \cite{FlFr2},\cite{Araki},\cite{singleton_physics}.  We shall obtain explicit solutions of this equation in the next section.
\section{Flato --- Fronsdal wave equation}
A fourth-order wave equation for singleton modes, 
\beq
\left( \Box_{CAdS_d} -\lambda \right)^2 F \,=\, 0,
\eeq{27}
was proposed by Flato and Fronsdal \cite{FlFr2} and used in a number of
publications. Here we obtain solutions of this equation and discuss their
properties.

We are looking for solutions of \reff{27} of the form
 $F = f(r)e^{-i\omega t}{\bf Y}({\bf \theta }) $.
Equation for $f(r)$ reads
\beq
Q^2 f\,=\, a_4(r)f^{IV} \,+\, a_3(r) f'''(r) \,+\, a_2(r)f''(r) \,+\, a_1(r) f'(r) \,+\, a_0(r)f(r) \,=\, 0,
\eeq{28}
where
\beq
a_4 \,=\, \cos^4 r,
\eeq{29}
\beq
a_3 \,=\, 2 (d-2) \cos^2 r \cot r - 4 \cos^3 r \sin r 
\eeq{30}
\begin{eqnarray}
a_2 &=& - 2 (d-2) \cos^2 r - 2 \frac{\lambda}{a^2} \cos^2 r 
- 2 \cos^4 r +2\omega^2\cos^4 r 
- 2 (d-2) \cot^2 r \nonumber \\ &+& (d-2)^2 \cot^2 r 
+ 2 \Lambda \cos^2 r \cot^2 r + 2 \cos^2 r \sin^2 r
\end{eqnarray}
\begin{eqnarray}
a_1 &=& -2 (d-2)\lambda \cot r + 2(d-2)\omega^2 \cos^2 r \cot r 
+
2(d-2)\cot^3 r - 4 \Lambda \cot^3 r \nonumber \\ &+&
 2 (d-2) \Lambda \cot^3 r  
- (d-2)^2 \cot r \csc^2 r - 4\omega^2 \cos^3 r \sin r
\end{eqnarray}
\begin{eqnarray}
a_0 &=& 2 \Lambda \cot^2 r\csc^2 r   - 2(d-2)\omega^2 \cos^2 r 
- 2 \frac{\lambda}{a^2}\omega^2 \cos^2 r - 2\omega^2 \cos^4 r 
 - 2\Lambda \frac{\lambda}{a^2}\cot^2 r \nonumber \\  &+&
2\Lambda \omega^2\cos^2 r \cot^2 r + \omega^4 \cos^4 r
+ 4\Lambda \cot^4 r + \Lambda^2\cot^4 r 
- 2 (d-2)\Lambda \cot^2 r \csc^2 r\nonumber \\  &+& 2 \omega^2 \cos^2 r \sin^2 r +
 \frac{\lambda^2}{a^4} 
\end{eqnarray}
where $\Lambda$ is the eigenvalue of $\triangle_{S^{d-2}}$ \reff{132}.
Roots of the indicial equation at $r=0$ are:
\beq
\alpha_1^0 \,=\,l, \; \; \;  \alpha_2^0 \,=\,l + 2, \; \; \; \alpha_3^0 \,=\,3 - d - l, \; \; \; \alpha_4^0 \,=\,5 - d -l .
\eeq{9001}
Roots of the indicial equation at $r=\pi /2$ are:
 \beq
\alpha_{1,2}^{\pi /2} = \frac{d-1}{2} \pm \sqrt{\left(  \frac{d-1}{2}\right)^2 +\frac{\lambda}{a^2}}.
\eeq{38}
\beq
\alpha_{3,4}^{\pi /2}  \,=\,
\alpha_{1,2}^{\pi /2} 
\eeq{39}
The generic solution with nonsingular behavior at the origin can be written as
\beq
f(r) \,=\, A \psi_1 (r) \,+\, B \psi_2 (r),
\eeq{391}
where
$\psi_1$ is given by \reff{22} and $\psi_2$ satisfies $Q\psi_2 = c \psi_1$, where $c$ is an arbitrary constant.  Solution nonsingular 
at $r=0$ can easily be constructed\footnote{The solution cannot be written in terms of hypergeometric function for the generic value of $E_0$.} 
\beq
\psi_2 \,=\, \sin^{l+2} r \cos^{E_0} r \left( 1 \,+\, c_1 \sin^2 r \,+\, c_2 \sin^4 r \,+\, \cdots \right),
\eeq{7688}
where
\beq
c_1 \,=\, \frac{(E_0 + l+1)^2 + 2l + d - \omega^2}{2(2l+d+1)},
\eeq{7843}
\beq
c_2 \,=\, \frac{\left(\omega^2 - (E_0+l+2)^2 - 2l - d - 3 \right)^2 + r_0}{8(2l+d+1)(2l+d+3)},
\eeq{2378}
\beq
r_0=-\frac{1}{3}\left( 16 E_0^2 + 8 E_0(6l+d+11) +12 l^2 -12l(d-9)-5(d-1)^2+128\right).
\eeq{45678}
When $E_0=(d-3)/2$ the above solution reduces to
\beq
\psi_2 \,=\,
  \sin^{l+2} r \cos^{\frac{d-3}{2}} r 
\; \ofo \left( \frac{l}{2}+\frac{d+1}{4} +\frac{\omega}{2},
 \frac{l}{2}+\frac{d+1}{4} -\frac{\omega}{2};
l+\frac{d+1}{2};\sin^2 r \right).
\eeq{392}
With the ``singleton frequency'', $\omega_k=(d-3)/2+l+2k$, $\psi_2$ gives two sets of modes,
\beq
F^{scalar}_{l,k} \,=\, \psi_2^{(k)} e^{-i (\frac{d+1}{2}+l+2k) t}{\bf Y(\theta )} =  e^{-i (\frac{d+1}{2}+l+2k) t} 
 \sin^{l+2} r \cos^{\frac{d-3}{2}} r  \, \, P_k^{(l+\frac{d-1}{2},
0)} \left( \cos 2r \right) {\bf Y(\theta )}
\eeq{392a}
and
\beq
\Psi^{(0)}_l =  e^{-i (\frac{d-3}{2}+l) t}\sin^{l+2}r  \cos^{\frac{d-3}{2}}r
 \; \Phi \left( \sin^2 r, 1,  l + \frac{d-1}{2}\right)
{\bf Y(\theta )},
\eeq{392b}
where
$k=0,1,\dots$, $\Phi (z,s,a)=\sum_{n=0}^{\infty}z^n/(n+a)^s$.
At the same time solution $\psi_1$ produces singleton modes  $F^{singleton}_l$ \reff{8199} 
and gauge modes  $F^{gauge}_{k,l}$ \reff{988a}.
The scalar, singleton and gauge modes form the Gupta - Bleuler triplet,
\beq
\left( \frac{d+1}{2}, scalar \right) \rightarrow \left( \frac{d-3}{2}, singleton \right) \rightarrow 
\left( \frac{d+1}{2}, gauge \right).
\eeq{392d}
which can be verified explicitly by 
acting on the states $|E_0,l,k>$ 
with the generators $M_i^{\pm}$\footnote{ 
For example, in $d=4$ one has $M_3^- |\frac{5}{2},0,0;\mbox{scal}>$
$=M_3^-(\sin^2 r e^{-i 5t/2}\cos^{1/2}r )$ $=
 2 i e^{-i3t/2} \sin r \cos^{1/2}r \cos 
\theta$ $ = 2 i | \frac{1}{2},1,0,\mbox{singleton}>$ and 
$M_3^+ |\frac{5}{2},0,0;\mbox{scal}>$ $= -i/2 e^{-i 7t/2}\cos^{1/2}r \cos \theta \sin r (5\cos 2r -1)$ $ =
 |\frac{5}{2},1,0;\mbox{gauge}>\oplus |\frac{5}{2},1,0;\mbox{scal}>$.}.

Let us examine the set $\{ \Psi_l^{(0)}\}$ more closely. Explicit expressions in odd/even dimension are:
\beq
\Psi_l^{(0)} = - \sin^{l+2}r \cos^{n-1}r \frac{l+n}{\sin^{2(l+n)}r}\left[ \log \left( \cos^2 r \right) +
\sum_{k=1}^{l+n-1} \frac{\sin^{2k}r}{k}\right],
\eeq{392e}
for $d=3,5,\dots 2n+1$,
\beq
\Psi_l^{(0)} =  \sin^{l+2}r \cos^{n-3/2}r \frac{2(l+n)-1}{\sin^{2(l+n)-1}r}
\left[ 
\mbox{arctanh} \left( \sin r \right) +
\sum_{k=1}^{l+n-1} \frac{\sin^{2k-1}r}{2k-1}\right].
\eeq{392f}
for $d=2,4,\dots 2n$.
In four dimensions, the two lowest states are given by
\beq
\Psi^{(0)}_0 = 3 e^{-it/2}\cos^{1/2}r \left[ \frac{\mbox{arctanh}(\sin r)}{\sin r} - 1 \right],
\eeq{add1}
\beq
\Psi^{(0)}_1 = 5 e^{-i 3t/2}\cos^{1/2}r \left[ \frac{\mbox{arctanh}(\sin r)}{\sin^2 r} - \frac{1}{\sin r} -  \frac{1}{3} \sin r  \right]\sin \theta .
\eeq{add2}
For $r<\pi/2$ they can be written in the form of power series
\beq
\Psi^{(0)}_0 = e^{-it/2}\sin^2 r \cos^{1/2}r \left( 1 + \frac{3}{5}\sin^2 r +
 \frac{3}{7}\sin^4 r + \cdots \right),
\eeq{add3}
\beq
\Psi^{(0)}_1 = e^{-i3t/2}\sin^3 r \cos^{1/2}r \left( 1 + \frac{5}{7}\sin^2 r +
 \frac{5}{9}\sin^4 r + \cdots \right)\sin \theta .
\eeq{add4}
While the energy raising operators act on \reff{add1}, \reff{add2}
 by the rule  
$M^+_i |\Psi^{(0)}_l> =  |\Psi^{(0)}_l>\oplus |\mbox{singleton}>$\footnote{For example, $M_3^+ \Psi^{(0)}_0 = 3 i/5  \Psi^{(0)}_1 - 2 i F_1^{singleton}$.}
the action of the energy lowering
operators gives the states with negative energy\footnote{$M_3^- \Psi^{(0)}_0 \sim
e^{it/2}$  }. 
The set  $\left\{ \Psi_l^{(0)} \right\}$ is therefore unacceptable and should be eliminated. This can be done by choosing 
\beq
\psi_2 \,=\, \frac{  \sin^{l+2} r \cos^{\frac{d-3}{2}} r}{\Gamma \left( \omega - \frac{d-3}{2}- l \right)}
\; \ofo \left( \frac{l}{2}+\frac{d+1}{4} +\frac{\omega}{2},
 \frac{l}{2}+\frac{d+1}{4} -\frac{\omega}{2};
l+\frac{d+1}{2};\sin^2 r \right).
\eeq{392x}
rather than \reff{392} 
as the second fundamental nonsingular at the origin solution of \reff{28}.

Finally,  the space of solutions of the fourth-order equation  \reff{28} is given by 
\beq
W= \left\{ F^{scalar} \right\}\oplus  \left\{ F^{singleton} \right\}\oplus  \left\{ F^{gauge} \right\}.
\eeq{392c}
with $F^{scalar}_{l,k}$, $F^{singleton}_l$ and $F^{gauge}_{l,k}$  given correspondingly by \reff{392a}, \reff{8199} and \reff{988a}.

Note that if the  vanishing flux boundary conditions are imposed, the only solution of the fourth-order equation is the Dirichlet set \reff{988a}. In that case the solution $\psi_2$ is trivial and thus the fourth-order equation gives nothing new in comparison with the second-order one.

\paragraph{Gupta - Bleuler triplet and the singleton two-point function}
According to the description of the singleton representation realized in the bulk of $CAdS$, the singleton two-point function satisfies Flato - Fronsdal wave equation \reff{27}. It is straightforward to generalize the four-dimensional result of \cite{FlFr2} to arbitrary $d$,
\beq
D_{FF}(Z) \,=\, Z^{-\frac{d-3}{2}}\,  \ofo \left( \frac{d-3 }{4},
 \frac{d-1}{4};
 1 ; \frac{1}{Z^2} \right).
\eeq{17add}
Having obtained explicit expressions for the modes we can now demonstrate 
that \reff{17add} admits decomposition of the form \reff{3tpf} with $\{\phi_i\}$ being the members of the Gupta - Bleuler triplet \reff{392c}. Indeed,
\beq
D_{FF} (Z) = 2^{\frac{d-3}{2}}\cos^{\frac{d-3}{2}}r e^{-i\frac{d-3}{2}t}
  \sum_{n=0}^{\infty} \frac{\left( \frac{d-3}{2}\right)_{n}}{n!}  e^{-i2nt}P_n^{(\frac{d-5}{2},0)}(\cos 2r ).
\eeq{11w}
Using \reff{90807a} it is not difficult to obtain the identity 
\beq
P_{k+1}^{(\alpha - 1 ,0)}\left( \cos 2r \right) \,=\, \frac{\alpha }{k+1}
 \cos^2 r 
P_{k}^{(\alpha ,1)}\left( \cos 2r \right) - \sin^2r P_{k}^{(\alpha +1 ,0)}\left( \cos 2r \right).
\eeq{90807aa}
Then
\begin{eqnarray}
D_{FF} (Z) &=& 2^{\frac{d-3}{2}}\cos^{\frac{d-3}{2}}r e^{-i\frac{d-3}{2}t} \nonumber \\ &+&
 2^{\frac{d-3}{2}}\cos^{\frac{d+1}{2}}r e^{-i\frac{d+1}{2}t}\left( \frac{d-3}{2}\right) \sum_{n=0}^{\infty} \frac{\left( \frac{d-3}{2}\right)_{n+1}}{n!(n+1)^2}  e^{-i2nt}P_n^{(\frac{d-3}{2},1)}(\cos 2r ) \nonumber \\
&-& 
 2^{\frac{d-3}{2}}\cos^{\frac{d-3}{2}}r \sin^2 r 
e^{-i\frac{d+1}{2}t} \sum_{n=0}^{\infty} \frac{\left( \frac{d-3}{2}\right)_{n+1}}{n!(n+1)^2}  e^{-i2nt}P_n^{(\frac{d-1}{2},0)}(\cos 2r ).
\end{eqnarray}
Three terms of this expression clearly correspond to the contribution of the singleton \reff{8199}, gauge  \reff{988a} and scalar \reff{392a} modes.

\appendix
\section{Transformation properties of $\ofo (a,b;c;z)$}\label{appendix_a}
In this appendix we record some useful formulae concerning the well-known transformation properties of the hypergeometric function.
 \begin{equation} \ofo (a,b;c;z) = (1-z)^{c-a-b}\ofo (c-a,c-b;c;z)
\label{9000}
\end{equation}
\begin{eqnarray}
\ofo (a,b;c;z) &=& \frac{\Gamma (c)\Gamma (c-a-b)}{\Gamma(c-a)\Gamma (c-b)}
\,  \, \ofo (a,b;a+b-c+1;1-z) \nonumber \\  &+&
(1-z)^{c-a-b} \frac{\Gamma (c)\Gamma (a+b-c)}{\Gamma(a)\Gamma (b)} \, \, \ofo (c-a,c-b;c-a-b+1
;1-z)
\label{999}
\end{eqnarray}
\begin{eqnarray}
\ofo (a,b,a+b;z) &=& \frac{\Gamma \left( a+b\right)}{\Gamma \left( 
a  \right) \Gamma \left( b  \right)}
\sum_{n=0}^{\infty} \frac{(a)_n(b)_n}{(n!)^2} \nonumber \\
&\cdot& \left(   2 \psi (n+1) - 
\psi (a+n) -\psi (b+n) -\log ( 1- z )\right) (1-z)^n ,
\label{a_3}
\end{eqnarray}
\begin{eqnarray}
\ofo (a,b,a+b+m;z) &=& \frac{\Gamma\left( m\right)\Gamma \left( a+b+m\right)}{\Gamma \left( 
a + m  \right) \Gamma \left( b + m \right)}\sum_{n=0}^{m-1}
 \frac{(a)_n(b)_n}{n! (1-m)_n}(1-z)^n \nonumber \\ &-&
 \frac{\Gamma \left( a+b+m\right)}{\Gamma \left( 
a  \right) \Gamma \left( b  \right)}(z-1)^m
\sum_{n=0}^{\infty} \frac{(a+m)_n(b+m)_n}{n!(m+n)!} \nonumber \\
&\cdot& [     - \psi (n+1) -\psi (n+m+1) +
\psi (a+n+m) +\psi (b+n +m)\nonumber \\  &+& \log ( 1- z )] (1-z)^n ,
\label{10004}
\end{eqnarray}
\begin{eqnarray}
\ofo (a,b,a+b-m;z) &=& \frac{\Gamma\left( m\right)\Gamma \left( a+b-m\right)}{\Gamma \left( 
a \right) \Gamma \left( b \right)}(1-z)^{-m}\sum_{n=0}^{m-1}
 \frac{(a-m)_n(b-m)_n}{n! (1-m)_n}(1-z)^n \nonumber \\ &-&
(-1)^m \frac{\Gamma \left( a+b-m\right)}{\Gamma \left( 
a -m  \right) \Gamma \left( b-m  \right)}
\sum_{n=0}^{\infty} \frac{(a)_n(b)_n}{n!(m+n)!} 
 [ - \psi (n+1) \nonumber \\ &-& \psi (n+m+1) +
\psi (a+n+m) 
 + \psi (b+n +m) +  
\log ( 1- z ) ]\nonumber \\ &\cdot& (1-z)^n ,
\label{10005}
\end{eqnarray}
\section{Two-dimensional case}\label{appendix_b}
Two-dimensional case is slightly different because in $d=2$ both 
solutions of the equation \reff{14}
are nonsingular at the origin.

The eigenvalue equation \reff{131} reads
\beq
\frac{\partial^2 F}{\partial t^2} \,-\, \frac{\partial^2 F}{\partial r^2} \,+\,\frac{\lambda}{a^2\cos^2 r}F \,=\,0, 
\eeq{19771}
where $\lambda =a^2 E_0(E_0-1)$. The solution is given by $F(r,t)=e^{-i\omega t}f(r)$, where
\beq
f(r) \,=\, A f_1(r) \,+\, B f_2 (r),
\eeq{19988}
\beq
f_1(r) \,=\,\cos^{E_0}r \ofo \left( \frac{E_0 -\omega}{2},  \frac{E_0 +\omega}{2};\frac{1}{2}; \sin^2 r \right),
\eeq{12098}
\beq
f_2(r) \,=\,\sin r \cos^{1-E_0}r \ofo \left( \frac{E_0 + 1 -\omega}{2},  \frac{E_0 + 1 + \omega}{2};\frac{3}{2}; \sin^2 r \right).
\eeq{12099}
Since the origin is a singular point of the limit-circle type we have to specify boundary conditions. Dirichlet condition $f(0)=0$ singles out solution \reff{12099}. Then the vanishing flux boundary conditions at $r=\pi /2$
give frequency quantization
$
\omega_k \,=\,2 \,-\,  E_0 \,+\, 2k,
$ $k=0,1,\dots$
and eigenfunctions
\beq
f^{DD}_k(r) \,=\, C_k \sin r \cos^{1-E_0}r P_k^{(\frac{1}{2},\frac{1}{2}-E_0)}(\cos 2r ),
\eeq{93949}
where
$
C_k^2 =  \Gamma \left( k - E_0 +2 \right) k! / \Gamma \left( k
 + 3/2 \right) \Gamma \left( k -E_0 +1/2\right)
$
(Dirichlet-Dirichlet set) or
$
\omega_k \,= E_0 \,+\, 2k \,+\,1 ,
$
\beq
f^{DN}_k(r) \,=\, C_k \sin r \cos^{E_0}r P_k^{(\frac{1}{2},E_0 - \frac{1}{2})}(\cos 2r ),
\eeq{93939}
where
$
C_k^2 = \Gamma \left( k + E_0 +! \right) k! / \Gamma \left( k
 + 3/2 \right) \Gamma \left( k + E_0 +1/2\right)
$
(Dirichlet-Neumann set), $k=0,1,\dots$.
Neumann condition $f'(0)=0$ singles out \reff{12098}. We obtain then
$
\omega_k \,=\,1 \,-\,  E_0 \,+\, 2k,
$
\beq
f^{ND}_k(r) \,=\, C_k  \cos^{1-E_0}r P_k^{(-\frac{1}{2},\frac{1}{2}-E_0)}(\cos 2r ),
\eeq{939439}
where
$
C_k^2 = \Gamma \left( k - E_0 +1 \right) k!/ \Gamma \left( k
 + 1/2 \right) \Gamma \left( k -E_0 +3/2\right)
$
(Neumann-Dirichlet set) or
$
\omega_k \,= E_0 \,+\, 2k,
$
\beq
f^{NN}_k(r) \,=\, C_k  \cos^{E_0}r P_k^{(-\frac{1}{2},E_0 - \frac{1}{2})}(\cos 2r ),
\eeq{96939}
where
$
C_k^2 = \Gamma \left( k + E_0  \right) k! / \Gamma \left( k
 + 1/2 \right) \Gamma \left( k + E_0 +1/2\right)
$
(Neumann-Neumann set), $k=0,1,\dots$.
These solutions correspond to the discrete series of UIRs of
 $\widetilde{SO}(1,2)$.
There are no positive energy singleton modes in this case
since  $E_0^{min}=-1/2<0$.

\section{Radial part of the Laplace equation on $CAdS_d$}
\label{appendix_c}
Unusual properties of the singleton modes can be further emphasized by
considering  the following example.
 One can redefine function $f(r)$ to absorb the measure          
in \reff{20}, $f(r) \,=\, \tan^{1-d/2}r \phi (r)$.
The square-integrability condition for $\phi$ is 
$
\int_0^{\pi /2}\phi^2dr <\infty
$
and equation \reff{14} reduces to the Schr\"{o}dinger-type one
$$
\left( - \frac{d^2}{dr^2} \,+\, V(r) \right) \phi \,=\, E,
$$
where
$E=\omega^2$,
$$
V(r) \,=\, \frac{l(l+d-3)}{\sin^2 r} \,+\,
 \frac{(d-2)(d-4)}{\sin^2 2r} \,-\,  
\frac{2-d - E_0(E_0 -d +1)}{\cos^2 r}.
$$
One finds that the ``energy'' $E$ is quantized, $E^{(k)}_l = (E_0 + l +2k)^2$, $k=0,1,\dots$.

\begin{figure}[h]
\centering
\epsfbox{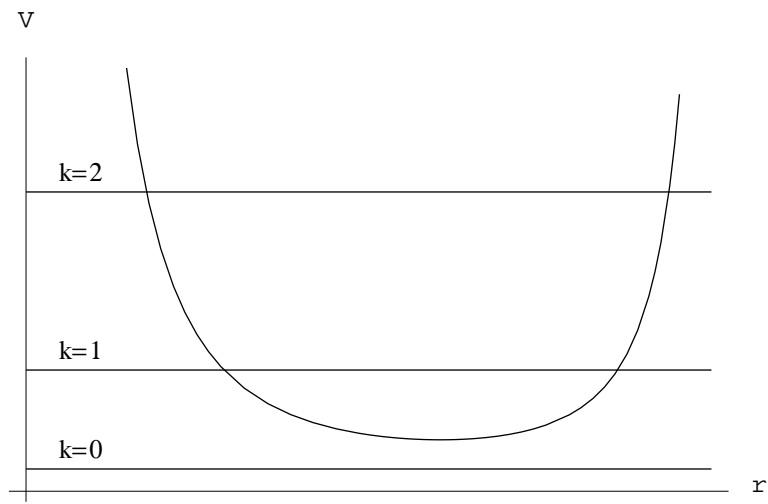}
\caption{Effective potential and ``energy'' levels for the singleton representation with $E_0=(d-3)/2$. Singleton wave function would have corresponded to the $k=0$ level.}
\label{fig2}
\end{figure}

For the singleton representation $E_0=(d-3)/2$. The set of formal solutions in this case consists of ``singleton modes'' ($k=0$)  and ``gauge modes''
$(k>1)$. 
Simple analysis shows that ``singleton modes''
correspond to the ``energy'' levels  $E^{(0)}_l$ which  lie entirely below the minimal level of the potential (see Figure~\reff{fig2}). These states cannot be given neither classical nor quantum-mechanical interpretation.

\vskip .2in
\noindent
\section*{Acknowledgments}
\vskip .1in
\noindent
I would like to thank Massimo Porrati, Martin Schaden and especially Christian  Fronsdal  
for useful and stimulating discussions, and Daniel Sternheimer for correspondence.

\end{document}